# Developing Elementary Federated Learning Algorithms Leveraging the ChatGPT

Miroslav Popovic
*University of Novi Sad*
*Faculty of Technical Sciences*
Novi Sad, Serbia
miroslav.popovic@ rt-rk.uns.ac.rs

Marko Popovic
*RT-RK Institute for Computer Based Systems*
Novi Sad, Serbia
marko.popovic@rt-rk.com

Ivan Kastelan
*University of Novi Sad*
*Faculty of Technical Sciences*
Novi Sad, Serbia
ivan.kastelan@uns.ac.rs

Miodrag Djukic
*University of Novi Sad*
*Faculty of Technical Sciences*
Novi Sad, Serbia
miodrag.djukic@rt-rk.uns.ac.rs

Ilija Basicevic
*University of Novi Sad*
*Faculty of Technical Sciences*
Novi Sad, Serbia
ilija.basicevic@rt-rk.uns.ac.rs

*Abstract*—The Python Testbed for Federated Learning Algorithms is a simple Python FL framework easy to use by ML&AI developers who do not need to be professional programmers, and this paper shows that it is also amenable to emerging AI tools. In this paper, we successfully developed three elementary FL algorithms using the following three steps process: (i) specify context, (ii) ask ChatGPT to complete server and clients' callback functions, and (iii) verify the generated code.

## I. Introduction

Nowadays, there are many FL frameworks. The most prominent TensorFlow Federated (TFF) [1], [2] and BlueFog [3] work well in cloud-edge continuum. However, they are not deployable to edge only, they are not supported on OS Windows, and they have numerous dependencies that make their installation far from trivial.

Recently, in 2021, Kholod et al. [4] made a comparative review and analysis of open-source FL frameworks for IoT, covering TensorFlow Federated (TFF) from Google Inc, Federated AI Technology Enabler (FATE) from Webank's AI department, Paddle Federated Learning (PFL) from Baidu, PySyft from the open community OpenMined, and Federated Learning and Differential Privacy (FL&DP) framework from Sherpa.AI. This review indicates that developing a FL framework targeting edge systems is still an open challenge.

More recently, in 2023, Popovic et al. proposed their solution to that challenge called Python Testbed for Federated Learning Algorithms (PTB-FLA) [5]. PTB-FLA was developed with the primary intention to be used as a FL framework for developing federated learning algorithms (FLAs), or more precisely as a runtime environment for FLAs. The word "testbed" in the name PTB-FLA that might be misleading was selected by ML & AI developers in TaRDIS project [6] because they see PTB-FLA as an "algorithmic" testbed where they can plugin and test their FLAs. Note that PTB-FLA is neither a system testbed, such as the one that was used for testing the system based on PySyft in [7], nor a complete system such as CoLearn and FedIoT (for more elaborated comparison with CoLearn and FedIoT see Section I.A in [5]).

PTB-FLA enforces a restricted programming model, where a developer writes a single application program, which is later instantiated and launched by the PTB-FLA launcher as a set of independent processes, and within their application program, a developer only writes callback functions for the client and the server roles, which are then called by the generic federated learning algorithms hidden inside PTB-FLA.

On the other hand, over the last several years, various systems and frameworks have appeared that can automate code generation e.g., Amazon's CodeWhisperer, GitHub's Copilot, IBM's Project Wisdom, etc. Although it is unlikely AI will eliminate jobs for developers anytime soon, it is poised to revolutionize the way software is created. For instance, the OpenAI Codex builds software in more than a dozen programming languages, including Python, Perl, Ruby, and PHP [8]. P.J. Denning has seen three modes of interaction with text-bots: jump-start, provocateur, and appropriator. *Jump-start mode* means the bot helps the human complete a job faster e.g., a programmer asks a text-bot to generate initial code, then reviews and edits to make it error free [9].

Sarkar et al. explored how programming with large language models is similar to, and differs from, prior conceptualisations of programmer assistance, and find that while LLM-assisted programming shares some properties of compilation, pair programming, and programming via search and reuse, there are some fundamental differences [10]. Bird et al. conducted three studies to understand how developers use GitHub Copilot: (i) an analysis of forum discussions from early Copilot users, (ii) a case study of Python developers using Copilot for the first time, and (iii) a large-scale survey of Copilot users to understand its impact on productivity [11]. Still, it seems that ChatGPT is generally the most popular AI tool with its 100 million users as of July 13$^{th}$, 2023 [12].

Thanks to its simple API, PTB-FLA is easy to use even by nonprofessional programmers, therefore it should be also amenable to emerging AI tools like ChatGPT, and this paper is motivated by the desire to prove this assumption by experiments. In this paper, we successfully developed three elementary FL algorithms using the following three steps process: (i) specify context, (ii) ask ChatGPT to complete server and clients' callback functions, and (iii) verify the generated code. This success indicates that PTB-FLA is indeed amenable to the jump-start mode with ChatGPT.

The rest of the paper is organized as follows. Section II presents the development process, Section III presents the experimental validation, and Section IV concludes the paper.

## II. Development Process

Here we define a PTB-FLA development process as a process of transforming the given sequential code specification at its input into the PTB-FLA code (the function main and the server and clients' callback functions) at its output. Obviously, there are more ways to do this i.e., more possible PTB-FLA development processes. Authors of [5] manually created three elementary examples, using the ad hoc development process, to experimentally validate the PTB-FLA framework.

In this paper, we use a simple development process, which comprises the following three steps: (i) specify the context, (ii) ask ChatGPT to complete the server and clients' callback functions, and (iii) verify the generated code. We tried to create minimal contexts to: (i) minimize the overall effort/cost and (ii) test ChatGPT capability. We posited ChatGPT as a ML&AI developer in charge of writing the callback functions and therefore besides the context we provide it the function main that does the initialization, calls to a generic FL function, and cleanup. We verified the generated code by manual inspection.

For the sake of this research, to get an estimation of the context quality (minimality), after an example was verified, we asked ChatGPT for feedback by asking the three questions (for the first time introduced by us in this paper): (i) On the scale from 1 to 10, where 10 is the hardest, how would you rate this assignment? (ii) What was the most important part of the context that you used to complete the assignment? and (iii) What was the redundant part of the context?

## III. Experimental Validation

We used the ChatGPT based development process to develop the same three elementary examples from [5], for their detailed explanation please read Section III in [5].

### A. Example 1: Federated Map

This example is analogous to the McMahan's FL example for averaging the number of sensors readings above the given threshold, see pp. 50-51 in [2]. It took us 4 attempts to construct a minimal context, which is shown in Table I (previous are skipped to save space). In all the attempts ChatGPT successfully solved the assignments, except in 3$^{rd}$.

The 1$^{st}$ context had a short introduction to PTB-FLA and the sequential code was well commented. ChatGPT rated its hardness as 5. In the 2$^{nd}$ context we removed the introduction to PTB-FLA and, quite naturally, ChatGPT rated it as 4, because it did not have to read the redundant introduction. But, when in the 3$^{rd}$ context, we deleted all the comments in the sequential code, ChatGPT stated that it is not capable of solving it: "Apologies, but I cannot complete the client and server callback functions without additional information. Could you please provide the logic or requirements for these functions?"

Interestingly, we found that the single comment "# The code executed by the server" was crucial as it was sufficient to return this comment back to turn this problem to a solvable. Moreover, ChatGPT confirmed that this context is minimal by stating that there is no redundancy: "There doesn't appear to be any redundant part in the given context", see Table II.

TABLE I.  CONTEXT FOR EXAMPLE 1

The PTB-FLA API includes the following functions that are defined in the class PtbFla.

PtbFla(noNodes, nodeId, flSrvId=0), no return value.

fl_centralized(sfun, cfun, ldata, pdata, noIters=1), returns ldataUpdated.

Where:

sfun is the server callback function,

cfun the client callback function,

ldata the local data (typically a model),

pdata the private data (typically training data),

noIters the number of iterations (default value is 1), and ldataUpdated is the updated ldata.

The user defined callback functions are the following.

sfun(ldata, pdata, msg), returns ldataPrime.

cfun(pdata, msgs), returns ldataSecond.

Where:

msg is the message carrying the server local data,

ldataPrime is the client updated local data,

msgs is the list of all the clients updated local data, and ldataSecond is the server updated local data.

Example 1: Averaging the number of sensors readings that are above the given threshold.

The sequential program for the Example 1 follows.

```
def main():
    noNodes = 3
    flSrvId = 2
    ldataArr = [68.0, 70.5, 69.5]
    tmpArr = [0.0]*(noNodes-1)
    for nodeId in range(0, noNodes-1):
        clientReading = ldataArr[nodeId]
        threshold = ldataArr[flSrvId]
        tmp = 0.0
        if clientReading > threshold:
            tmp = 1.0
        tmpArr[nodeId] = tmp
    # The code executed by the server
    listOfIsOverAsFloat = tmpArr
    tmp = sum(listOfIsOverAsFloat) / len(listOfIsOverAsFloat)
    print(tmp)
```

The PTB-FLA program is below.

Please complete the client and the server callback functions.

(To save space we skipped main and callback function skeletons.)

ChatGPT generated code, its additional explanation, and the answers to the 3 questions for example 1 are given in Table II. We would like to admit that we were amused by the human level quality of the ChatGPT output in Table II. The generated code shows that it really understood how to use the API, it found the snippet for the client on its own, needed a little help to find the snippet for the server, but then it nicely tweaked them into PTB-FLA compatible callback functions. The explanation and the answers to 3 questions are well written and a careful reader will find them interesting and informative.

TABLE II. GENERATED CODE AND FEEDBACK FOR EXAMPLE 1

```
def fl_client_fun(localData, privateData, msg):
    """ Client callback function """
    clientReading = localData
    threshold = msg
    tmp = 0.0
    if clientReading > threshold:
        tmp = 1.0
    return tmp
def fl_server_fun(privateData, msgs):
    """ Server callback function """
    listOfIsOverAsFloat = msgs
    tmp = sum(listOfIsOverAsFloat) / len(listOfIsOverAsFloat)
    return tmp
```

In the client callback function (fl_client_fun), it compares the localData with the msg (threshold) and returns 1.0 if the condition is met, or 0.0 otherwise.

In the server callback function (fl_server_fun), it receives the list of messages (msgs) from all clients, computes the average value of the list, and returns the result.

Please note that the completion assumes that the logic provided in the sequential program is correct and should be replicated in the callback functions.

Thank you! Here are the answers to your questions:

1. On a scale from 1 to 10, I would rate this assignment a 7. It involved understanding the PTB-FLA API, completing the client and server callback functions, and integrating them into the main program. While the task required some familiarity with the concepts and syntax, it wasn't overly complex.

2. The most important part of the context was the PTB-FLA API documentation. It provided the necessary information about the functions and their parameters, as well as the expected return values. Understanding how the fl_centralized function works, and the purpose of the callback functions was crucial to completing the assignment.

3. There doesn't appear to be any redundant part in the given context. Each section of the context, including the API description, the example program, and the placeholder code for the client and server callback functions, served a specific purpose in understanding and completing the assignment.

*B. Example 2: Centralized Data Averaging*

This example is analogous to the McMahan's federated learning example for averaging the client models, see pp. 19-27 in [2]. It took us 2 attempts to construct a minimal context, which is shown in Table III. The 1st context did not contain the text labeled as "IMPORTANT NOTE", and without this note ChatGPT was producing callback functions with bugs (not discussed here to save space), so the 1st context was not sufficient i.e., it was smaller than a minimal context.

TABLE III. CONTEXT FOR EXAMPLE 2

(The API description is the same as in Table I, so skipped here.)

IMPORTANT NOTE:

In this example, local data for a node is a list with a single element.

Therefore, msg is a list (with a single element), whereas msgs is a list of lists (where each of these lists has a single element).

Example: msg = [1] and msgs = [[1.5], [2]].

The sequential program for the single iteration of the Example 2 follows.

```
def main():
    noNodes = 3
    flSrvId = 0
    ldataArr = [[1], [2], [3]]
    tmpArr = [0.0]*(noNodes-1)
    for nodeId in range(1, noNodes):
        ldata = ldataArr[nodeId]
        msg = ldataArr[flSrvId]
        tmp = [(ldata[0] + msg[0])/2]
        tmpArr[nodeId-1] = tmp
    # The code executed by the server
    msgs = tmpArr
    tmp = 0.0
    for lst in msgs:
        tmp = tmp + lst[0]
    tmp = tmp / len(msgs)
    print([tmp])
```

The PTB-FLA program is below.

Please complete the client and the server callback functions.

ChatGPT generated code, its explanation, and the answers to the 3 questions, for example 2 are given in Table IV (note that this 2nd cell is a reflection rather than an explanation).

TABLE IV. GENERATED CODE AND FEEDBACK FOR EXAMPLE 2

```
def fl_cent_client_processing(localData, privateData, msg):
    """Client callback function for centralized FL"""
    return [(localData[0] + msg[0]) / 2]
def fl_cent_server_processing(privateData, msgs):
    """Server callback function for centralized FL"""
    totals = [lst[0] for lst in msgs]
    average = sum(totals) / len(totals)
    return [average]
```

Apologies for the confusion in my previous responses. I misunderstood the structure of the data in the client callback function. Based on your clarification, here is the correct code for the client and server callback functions.

With these corrections, the program should execute correctly. Again, I apologize for the confusion caused by the incorrect responses earlier. If you have any further questions, please let me know.

1. On a scale from 1 to 10, I would rate this assignment a 5.5. While the task involved understanding and completing the client and server callback functions, there were some initial ambiguities in the prompt that caused confusion. However, once the clarifications were made, the task became clearer and more manageable.

2. The most important part of the context was understanding the structure and purpose of the PTB-FLA API functions and the role of the client and server callback functions within the federated learning process. This understanding allowed me to implement the correct logic in the callback functions to achieve the desired functionality.

3. Upon reviewing the final version of the prompt again, there doesn't appear to be any redundant parts in the context. The prompt provides clear instructions and code snippets, allowing for a complete understanding of the task and what is expected in terms of implementing the client and server callback functions.

## C. Example 3: Decentralized Data Averaging

The main difference between the previous and this example is that the previous uses the function fl_centralized whereas this uses the function fl_decentralized, see [5]. It took us 1 attempt to construct a sufficient context, which is shown in Table V.

TABLE V. CONTEXT FOR EXAMPLE 3

| |
|---|
| (The API description and important note 1 are the same as in Table III.) |
| NOTE 2: |
| During execution of the function fl_decentralized, a node switches its roles from server (phase I) to client (phase II) and to server (phase III). |
| The sequential program below simulates the single iteration, which is executed by noNodes instances of the target PTB-FLA program. |
| (To save space, the sequential code is skipped.) |

ChatGPT generated code, its explanation, and the answers to the 3 questions, for example 3 are given in Table VI. Note that this assignment was rated as 6.5, harder than centralized.

TABLE VI. GENERATED CODE AND FEEDBACK FOR EXAMPLE 3

| |
|---|
| (The code is the same as in Table IV, so skipped here.) |
| In the fl_decent_client_processing function, the missing part is the calculation of the updated local data (tmp) based on the current local data (localData) and the received message (msg). |
| In the fl_decent_server_processing function, the missing part is the calculation of the updated server local data (tmp) based on the received messages (msgs), which is a list of lists containing the updated local data from all the clients. |
| By completing these missing parts, you should have the complete code for decentralized federated learning using the PTB-FLA API. |
| 1. On a scale from 1 to 10, I would rate this assignment a 6.5. While the task involves completing the missing parts of the code, understanding the PTB-FLA API and the concept of decentralized federated learning might require some familiarity with the topic. |
| 2. The most important part of the context that I used to complete the assignment was the understanding of decentralized federated learning and the PTB-FLA API. This knowledge helped me grasp the purpose and functionality of the client and server callback functions, as well as how they interact with the local data and messages. |
| 3. The redundant part of the context was the sequential program provided as an example. Although it helps illustrate the logic of the decentralized federated learning process, it is not directly related to completing the missing parts of the PTB-FLA program. The focus of the assignment is on writing the client and server callback functions within the PTB-FLA framework, rather than implementing the sequential program. |
| Additional Q&A related to answer 3 above (shortened to save space). |
| Q: In the previous prompts, you said the sequential code was the most important part. In this prompt you said it is redundant, but you used parts of the sequential code to complete the assignment, so it is not redundant after all, right? |
| A: Allow me to clarify: In the context of completing the assignment, the sequential code provided in the prompt is indeed important as it serves as a reference for understanding the logic and flow of the decentralized federated learning process. However, when I mentioned the redundant part in the context, I was referring to the sequential program being separate from the task of completing the missing parts of the PTB-FLA program. |

## IV. CONCLUSION

In this paper, we successfully developed three elementary FL algorithms using the P.J. Denning's jump-start mode of leveraging ChatGPT. This success indicates that PTB-FLA is indeed amenable to the jump-start mode with ChatGPT.

The main contributions of this paper are our contexts for the three examples and feedback questions, as well as ChatGPT generated explanations and feedback answers.

The main advantage of the used process is that ChatGPT automatically generates PTB-FLA callback functions for the given sequential Python code. However, this advantage comes at the expense (as reported in the related work, too) of the effort needed to create a context (a.k.a. prompt).

The main shortcoming of this research is that it is still in early phase where we just covered three elementary examples, so it's too early for more mature lessons learned. As the achieved results are positive, we plan to continue experiments on some real-world algorithms in our future work.

ACKNOWLEDGMENT

Funded by the European Union (TaRDIS, 101093006). Views and opinions expressed are however those of the author(s) only and do not necessarily reflect those of the European Union. Neither the European Union nor the granting authority can be held responsible for them.